\documentclass[11pt]{article}
\usepackage{latexsym}
\usepackage{amssymb}
\usepackage{epsf}
\usepackage{amsmath}
\usepackage{graphicx}
\usepackage{multirow}

\newcommand{\lsim}{\lesssim}

\begin{document}
\pagestyle{empty}

\begin{flushright}
YITP-13-37 \\ 
\end{flushright}

\vspace{3cm}

\begin{center}

{\bf\LARGE Minimal Fine Tuning \\ \vskip.5cm in Supersymmetric Higgs Inflation}
\\

\vspace*{1.5cm}
{\large 
Takumi Imai$^1$ and Izawa K.-I.$^{1,2}$
} \\
\vspace*{0.5cm}

$^1${\it Yukawa Institute for Theoretical Physics, Kyoto University, Kyoto
 606-8502, Japan}\\
\vspace*{0.3cm}
$^2${\it Kavli Institute for the Physics and Mathematics of the Universe (WPI),\\
 University of Tokyo, Kashiwa 277-8583, Japan}\\
\vspace*{0.5cm}

\end{center}

\vspace*{1.0cm}

\begin{abstract}
{
\pagestyle{plain}
We investigate characteristic features of realistic parameter choice
for primordial inflation with supersymmetric Higgs inflaton
as an example of particle physics inflation model. 
We discuss constraints from observational results
and analyze the degree of fine tuning needed to induce slow-roll inflation
for wide range of soft supersymmetry breaking scale.
The observed amplitude of density fluctuations implies that
the minimal fine tuning for the combined electroweak scale and inflaton flatness
predicts the spectral index of $n_{\rm s}=0.950-0.965$,
which includes the central value from observational data.
}
\end{abstract} 

\newpage
\baselineskip=18pt
\setcounter{page}{2}
\pagestyle{plain}
\baselineskip=18pt
\pagestyle{plain}

\setcounter{footnote}{0}

\section{Introduction}

Recent experimental information such as the Planck satellite results
\cite{Ade:2013uln}
provides cosmological parameters with increasing accuracy,
which enables us to perform detailed numerical examination
of realistic model parameters in various candidate inflation models
\cite{Martin:2013tda}.
One of the motivations to investigate numerical aspects of
concrete inflation models is to see how the model parameters are
chosen or tuned to realize appropriate slow-roll inflation.

In this paper, among various models of inflation, we adopt
a supersymmetric Higgs inflation model
\cite{Chatterjee:2011qr}
to investigate its quantitative aspects
as a simple example of particle physics model of inflation.
In addition to the parameter tuning for inflation,
such a particle physics model of inflation may be relevant for
electroweak hierarchy tuning. We are also confronted with no discovery
of superpartners so far.
Hence we do not restrict ourselves
to the case of weak scale supersymmetry (SUSY) in contrast to the conventional analysis.

As for the supersymmetric Higgs degrees of freedom,
the quartic contributions to the scalar potential due to gauge interactions
are identically zero
for specific directions called D-flat directions in the field space of Higgs scalars.
Along these flat directions, the scalar potential is determined by the
soft SUSY breaking terms and higher order non-renormalizable operators
appearing in the effective low energy theory. For suitable parameter
choices, such a scalar potential can realize successful primordial
inflation. 

The rest of the paper is organized as follows. In the next section,
we present the setup of supersymmetric Higgs inflation model,
whose D-flat direction is modified to have a nearly flat inflection point.
We identify the model parameters which are fine tuned to realize
slow-roll inflation. 
In section 3, the observational constraints on temperature fluctuations
of cosmic microwave background radiation are taken into account
to obtain its index of spectral tilt in the supersymmetric Higgs
inflation model as depicted in tables 1$-$3. In section 4, we consider
inflationary parameter tuning to induce slow-roll inflation, the observed
amplitude of density fluctuations, or eternal inflation in the present
setup. The final section concludes the paper.

\section{The inflaton potential}

We first recapitulate the supersymmetric Higgs inflation
model considered by Chatterjee and Mazumdar \cite{Chatterjee:2011qr}.
The Higgs fields are those of the minimal supersymmetric standard model (MSSM)
with up-type and down-type ones denoted by $H_u$ and $H_d$, respectively.

\subsection{D-flat direction}

Let us parametrize the D-flat direction of the Higgs potential as follows:
\begin{eqnarray}
H_u = (\frac{1}{\sqrt{2}}\Phi,0),~~~~
H_d = (0,\frac{1}{\sqrt{2}}\Phi),
\end{eqnarray}
where $\Phi$ is a complex scalar field. 
The quartic terms in the potential would vanish along this direction
in the case of genuine MSSM.

The potential can be modified by non-renormalizable terms 
which originate from the Higgs superpotential
\begin{eqnarray}
{W}=\mu \mathcal{H}_u\cdot \mathcal{H}_d+\sum_{k\geq2}\frac{\lambda_k}{k}\frac{(\mathcal{H}_u\cdot\mathcal{H}_d)^k}{M_{\rm P}^{2k-3}},
\end{eqnarray} 
where $M_P$ denotes the reduced Planck mass $\simeq 2.4\times 10^{18}$GeV
and $\mathcal{H}_u$ and $\mathcal{H}_d$ are superfields
corresponding to $H_u$ and $H_d$.
Hereafter, we only keep the lowest additional term with $k=2$
in this paper.

Taking into account soft SUSY breaking terms, we have the scalar potential along the D-flat direction as
\begin{eqnarray}
{V}(\phi,\theta) = \frac{1}{2}m^2(\theta)\phi^2-\frac{\lambda_2 \mu}{4M_{\rm P}} \cos(2\theta)\phi^4+\frac{{\lambda_2}^2}{32M_{\rm P}^2}\phi^6,
\end{eqnarray} 
where $\phi$ and $\theta$ denote the radial and angular components of the field $\Phi=\frac{1}{\sqrt{2}}\phi e^{i\theta}$ and 
\begin{eqnarray}
m^2(\theta) &=& \frac{1}{2}(2\mu^2+m^2_{H_u}+m^2_{H_d}-2b\cos{2\theta}).
\end{eqnarray}
The soft parameters $m^2_{H_u}$, $m^2_{H_d}$, and $b$ are coefficients
of Higgs soft quadratic terms in the potential.
Here, we have taken all the parameters to be positive.

The potential is minimized along the angular direction when $\theta=0$
with the effective $\theta$ mass above the Hubble mass during inflation.
Then it is given by 
\begin{eqnarray}
V(\phi) \equiv {V}(\phi,0)= \frac{1}{2}m^2_0\phi^2-\frac{\lambda_2 \mu}{4M_{\rm P}} \phi^4+\frac{{\lambda_2}^2}{32M_{\rm P}^2}\phi^6, \label{potential}
\end{eqnarray} 
where $m_0 \equiv m(\theta=0)$. This $m_0$ may be regarded
as a typical soft SUSY breaking scale so that the hierarchy
between $m_0$ and the Z boson mass $m_Z$ indicates the electroweak
fine tuning.

\subsection{Inflection point and slow-roll parameters}

For the primordial Higgs inflation to occur successfully,
the potential at least has to possess some region where the slow-roll
conditions are satisfied.
Actually, such a flat region is realized if the potential has
an inflection point almost like a saddle point.

The potential \eqref{potential} has a saddle point when the parameter
relation $3m_0^2=4\mu^2$ holds.
Hence we define a characteristic parameter $\alpha$ as
\begin{eqnarray}
3m^2_0=4\mu^2(1+8\alpha^2),
\end{eqnarray} 
which parameterizes the tuning required for slow-roll inflation,
reminiscent of electroweak scale tuning in MSSM.
We restrict ourselves to the regime $\alpha^2>0$, for which
the potential is monotonic around the inflection point.

If this parameter is fine-tuned as $\alpha^2 \ll 1$, successful slow-roll inflation can occur near the inflection point $\phi=\phi_0$:
\begin{eqnarray}
\phi_0 &=&  \Big(\frac{4 M_P}{\sqrt{3}\lambda_2}m_0\Big)^{\frac{1}{2}}(1-\alpha^2)+\mathcal{O}(\alpha^4).
\end{eqnarray} 
We assume $\phi_0 \ll M_P$, that is, $m_0 \ll \lambda_2 M_P$.
Around the inflection point $\phi_0$, the potential can be expanded as
\begin{eqnarray}
V(\phi) &=& V_0+\beta_1(\phi-\phi_0)+\frac{1}{6}\beta_3(\phi-\phi_0)^3+ \cdots,
\end{eqnarray}
where
\begin{eqnarray}
V_0 &=& V(\phi_0) = \frac{1}{6}m^2_0\phi^2_0+\mathcal{O}(\alpha^2), \\
\beta_1
&=& 8\alpha^2 m_0^2 \phi_0+\mathcal{O}(\alpha^4), \\
\beta_3
&=& 8 \frac{m_0^2}{\phi_0}+\mathcal{O}(\alpha^2).
\end{eqnarray} 
The ellipsis represents higher order terms, whose effects
during inflation we neglect in the following analysis.

The slow-roll parameters are given by
\begin{eqnarray}
\epsilon(\phi) &\equiv& \frac{M_P^2}{2}\Big(\frac{V'}{V}\Big)^2 \simeq \frac{M_P^2}{2V_0^2}(\beta_1+\frac{\beta_3}{2}(\phi-\phi_0)^2)^2, \label{eps} \\
\eta(\phi) &\equiv& M_P^2\frac{V''}{V} \simeq M_P^2\frac{\beta_3}{V_0}(\phi-\phi_0). \label{et}
\end{eqnarray} 
The absolute values of these parameters are smaller than one
during inflation.

The inflation ends when the slow-roll parameter $\eta$ reaches one so that
the end point $\phi=\phi_{\rm end}$ of slow-roll inflation is given by
\begin{eqnarray}
\phi_0-\phi_{\rm end} \simeq \Big(\frac{\phi^3_0}{48M^2_{\rm P}}\Big).
\end{eqnarray} 
This should be small enough to neglect higher order terms in the potential.

\section{Observational constraints}

The parameters which characterize the primordial inflation model
are constrained by observational data.
In particular, we have restrictions on
the amplitude of density fluctuations $\sqrt{A_{\rm s}}
\simeq 4.69\times 10^{-5}$ and its spectral tilt
parameterized by the spectral index $n_s \simeq 0.96$
with narrowing uncertainties \cite{Ade:2013uln}.

For the supersymmetric Higgs inflation model described in the previous
section, these quantities are obtained
\cite{Bueno Sanchez:2006xk, Allahverdi:2006we}
by means of inflaton equations of motion as
\begin{eqnarray}
\sqrt{A_{\rm s}}
&\simeq& \frac{1}{72\sqrt{6}\pi}\frac{1}{X^2\lambda_2}\sin^2[6\sqrt{6}\mathcal{N}X\lambda_2]\label{amplitude of density fluctuation}, \\
n_{\rm s}
&\simeq& 1-24\sqrt{6}X\lambda_2 \cot[6\sqrt{6}\mathcal{N}X\lambda_2], \label{spectral index}
\end{eqnarray} 
where $X \equiv \alpha\big(\frac{M_P}{m_0}\big)$ and $\mathcal{N}$ is the
number of e-foldings of the present horizon after horizon exit during
inflation. Note that the $X$ indicates the combined degree of fine
tuning for inflationary potential
flatness and hierarchy between the soft SUSY breaking scale $m_0$ and the
electroweak scale $m_Z$.
For a fixed value of $X$, the $\alpha$ tuning can be compensated
by the $M_P/m_0$ tuning or the $m_Z/m_0$ tuning up to the constant factor $M_P/m_Z$.

It turns out through numerical estimates that $X\lsim 10^4$ and
$\lambda_2\sim10^{-7}$ are needed to produce
the right amount of density fluctuations $\sqrt{A_{\rm s}}
\simeq 4.69\times 10^{-5}$.
We list appropriate sample
parameters sets and the corresponding theoretically obtained $n_{\rm s}$
in tables 1, 2, and 3, which
show the cases of $\mathcal{N}=40, 50$, and $60$
under $\sqrt{A_{\rm s}}= 4.69\times 10^{-5}$.
The uncertainty of $\sqrt{A_{\rm s}}$ is negligible compared to
those of $\mathcal{N}$ and $n_{\rm s}$ in our analysis.
For each value of $X$,
there are two values of $\lambda_2$ which satisfy the constraints from
the observed amplitude of density fluctuations.
They are denoted by
$\lambda_2^{(1)}$ and $\lambda_2^{(2)}$ and the corresponding values of
$n_{\rm s}$ are obtained as $n_{\rm s}^{(1)}$ and $n_{\rm s}^{(2)}$.
The parenthesized values in the tables are out of assumed 
slow-roll inflation and written just for completeness.
The $X$ takes the maximal values (see the next section) for the limiting case of
$\lambda_2^{(1)}=\lambda_2^{(2)}$ and $n_{\rm s}^{(1)} = n_{\rm s}^{(2)}$,
which is denoted in the captions to the tables.

\section{Inflationary parameter tuning}

Based on our numerical analysis, we now consider possible criteria
for inflationary parameter tuning.

The obvious minimum requirement is that inflationary phase is at least
present.
The equations \eqref{eps} and \eqref{et} for slow-roll parameters
yield the indispensable condition for slow-roll inflation to occur as
$\epsilon(\phi_0 \pm \Delta\phi_{\rm q}) \lsim 1$ and $\eta(\phi_0 \pm \Delta\phi_{\rm q}) \lsim 1$, which amount to
\begin{eqnarray}
X^4 \lsim \frac{1}{288\sqrt{3}\lambda_2}\frac{M_{\rm P}^3}{m_0^3}, \qquad
\lambda_2 \lsim \frac{\pi}{\sqrt{6}}
\label{large alpha}.
\end{eqnarray}
Here, $\Delta\phi_{\rm q}$ denotes quantum variance
$\Delta\phi_{\rm q} \simeq H/2\pi$ in the field $\phi$ during Hubble time $H^{-1}$.
Thus, the smaller $m_0/M_P$ is, the less the necessary fine tuning for
$X$ is. 

As a realistic primordial inflation, sizable density fluctuations
should be produced during inflationary phase.
Let us reflect on the expressions \eqref{amplitude of density fluctuation}
and \eqref{spectral index} for density fluctuations and spectral index. 
For a fixed value of the amplitude $\sqrt A_{\rm s}$,
the allowed value of $X$ has an upper bound, since the function
$\sin^2x/x$ in Eq.\eqref{amplitude of density fluctuation}
has an upper bound, which is about 0.725 at $x=1.166$.
The observed amplitude of density fluctuations implies that
the maximal values of $X$ are $1.641\times10^4-2.461\times10^4$ for
$\mathcal{N}=40-60$, resulting in $n_{\rm s}=0.950-0.965$,
which captures the central value of the observed spectral index.
This maximal value of $X$ is none other than the minimal fine tuning
in the present supersymmetric Higgs inflation.
Note that the advantage of small $m_0/M_P$ in fine tuning is absent here
contrary to the case of mere tuning for slow-roll inflation to occur.

As a possible further fine tuning, we finally consider the parameter
tuning to realize eternal inflation.
If the fine-tuned parameter $\alpha$ is exceedingly small, the first derivative of the
potential near the inflection point $\phi_0$ is extremely tiny. In such
a case, when $\phi$ is very close to $\phi_0$, quantum effects dominate
field fluctuations and keep the system in a de Sitter
background effectively. This eternal inflation regime exists if quantum variance
$\Delta\phi_{\rm q} \simeq H/2\pi$ in the field $\phi$ during Hubble time $H^{-1}$ is larger than
the corresponding classical change $\Delta\phi_{\rm c}\simeq |-V'/3H^2|$.

In the present model, this condition turns out to be 
\begin{eqnarray}
(\phi-\phi_0)^2 \lsim \frac{1}{144\sqrt{2}\pi}\frac{m_0}{M^3_{\rm P}}\phi_0^4-2\alpha^2\phi_0^2.
\end{eqnarray}
For the field to be randomly kept inside the eternal inflation regime,  the right-hand
side of the inequality have to be larger than the inflaton quantum
variance squared during the Hubble time. This requires
\begin{eqnarray}
X^2 \lsim \frac{1}{72\pi}\Big(\frac{1}{\sqrt{6}\lambda_2}-\frac{1}{2\pi}\Big)\label{small alpha}.
\end{eqnarray}
Namely, if $\alpha$ is small enough to satisfy this inequality,
the eternal inflation regime exists
near the inflection point. For $\lambda_2 \sim 10^{-7}$, this condition
turns out to be $X \lsim 10^2$, which yields $n_{\rm s}=0.900-0.933$
for $\mathcal{N}=40-60$.

\section{Conclusion}

We have investigated inflationary parameter tuning
in the supersymmetric Higgs inflation model as an example
of particle physics model of primordial inflation.
The observed tilt of the spectral index implies that
the tuning is minimal to realize the sizable amplitude
of density fluctuations.

That is, under the observed amplitude of density fluctuations,
the maximal values of the simultaneous fine tuning parameter
$X = \alpha\big(\frac{m_Z}{m_0}\big)\big(\frac{M_P}{m_Z}\big)$
for the electroweak scale and inflaton flatness 
is given by $X=1.641\times10^4-2.461\times10^4$ for e-fold numbers
$\mathcal{N}=40-60$, resulting in the spectral index $n_{\rm s}=0.950-0.965$,
which includes the central value of the observed spectral index.
The advantage of small $m_0/m_Z$ in electroweak fine tuning is
compensated by the inflaton flatness tuning to realize sizable density fluctuations.
Namely, the weak scale SUSY does not ameliorate the total degree of fine
tuning including electroweak hierarchy
in the present setup once we take into account the inflationary fine tuning.

We performed a simple case study of supersymmetric Higgs inflation model
in this paper.
Inflationary parameter tuning may be intriguing to explore
in various models of inflation
\cite{Martin:2013tda},
by one of which we suspect the primordial inflation of our universe
is well described.
Future observations will further constrain realistic inflation models
and enable us to make even more detailed examination
on parameter choices thereof. We hope that this serves to reveal
fundamental structures to determine model parameters in particle physics.

\newpage

\section*{Acknowledgments}

This work was supported by
World Premier International Research Center Initiative
(WPI Initiative), MEXT, Japan.

\vskip1.9cm

\begin{table}[htb]
\begin{center}
\scalebox{1}
{\renewcommand{\arraystretch}{1.5}
\begin{tabular}{|c||c|c||c|c|} \hline
~~$X$~~& ~~$\lambda_2^{(1)}$~~&~~$n^{(1)}_{\rm s}$~~& ~~$\lambda_2^{(2)}$~~&~~$n^{(2)}_{\rm s}$~~ \\ \hline
$1.6\times10^4$&$1.054\times10^{-7}$&$0.935$&$1.429\times10^{-7}$&$0.969$ \\ \hline
$1.4\times10^4$&$0.908\times10^{-7}$&$0.919$&$1.961\times10^{-7}$&$1.007$ \\ \hline
$1.2\times10^4$&$0.847\times10^{-7}$&$0.912$&$2.543\times10^{-7}$&$1.041$ \\ \hline
$1.0\times10^4$&$0.810\times10^{-7}$&$0.908$&$3.319\times10^{-7}$&$1.078$ \\ \hline
$1.0\times10^3$&$0.751\times10^{-7}$&$0.900$&$4.734\times10^{-6}$&$1.744$ \\ \hline
$1.0\times10^2$&$0.751\times10^{-7}$&$0.900$&$(5.146\times10^{-5})$&($3.600$) \\ \hline
$1.0\times10^1$&$0.751\times10^{-7}$&$0.900$&$(5.280\times10^{-4})$&($9.380$) \\ \hline
\end{tabular}
}
\end{center}
\caption{Sample parameters sets for the $\mathcal{N}=40$ case. The
 maximal value of $X$ to achieve $\sqrt{A_{\rm s}} = 4.69\times 10^{-5}$ is $X=1.641\times 10^4$ with $\lambda_2=1.207\times 10^{-7}$ and $n_{\rm s}=0.950$. The parenthesized values are written just for completeness.}
\label{40}
\end{table}

\begin{table}[htb]
\begin{center}
\scalebox{1}
{\renewcommand{\arraystretch}{1.5}
\begin{tabular}{|c||c|c||c|c|} \hline
~~$X$~~& ~~$\lambda_2^{(1)}$~~&~~$n^{(1)}_{\rm s}$~~& ~~$\lambda_2^{(2)}$~~&~~$n^{(2)}_{\rm s}$~~ \\ \hline
$2.0\times10^4$&$0.675\times10^{-7}$&$0.948$&$0.914\times10^{-7}$&$0.975$ \\ \hline
$1.8\times10^4$&$0.592\times10^{-7}$&$0.937$&$1.189\times10^{-7}$&$1.000$ \\ \hline
$1.6\times10^4$&$0.555\times10^{-7}$&$0.932$&$1.468\times10^{-7}$&$1.022$ \\ \hline
$1.4\times10^4$&$0.531\times10^{-7}$&$0.928$&$1.807\times10^{-7}$&$1.044$ \\ \hline
$1.2\times10^4$&$0.515\times10^{-7}$&$0.926$&$2.247\times10^{-7}$&$1.069$ \\ \hline
$1.0\times10^4$&$0.503\times10^{-7}$&$0.924$&$2.861\times10^{-7}$&$1.099$ \\ \hline
$1.0\times10^3$&$0.481\times10^{-7}$&$0.920$&$3.837\times10^{-6}$&$1.678$ \\ \hline
$1.0\times10^2$&$0.481\times10^{-7}$&$0.920$&$(4.133\times10^{-5})$&($3.334$) \\ \hline
$1.0\times10^1$&$0.481\times10^{-7}$&$0.920$&$(4.229\times10^{-4})$&($8.501$) \\ \hline
\end{tabular}
}
\end{center}
\caption{Sample parameters sets for the $\mathcal{N}=50$ case. The
 maximal value of $X$ to achieve $\sqrt{A_{\rm s}} = 4.69\times 10^{-5}$ is $X=2.051\times 10^4$ with $\lambda_2=0.774\times 10^{-7}$ and $n_{\rm s}=0.960$. The parenthesized values are written just for completeness.}
\label{50}
\end{table}

\begin{table}[htb]
\begin{center}
\scalebox{1}
{\renewcommand{\arraystretch}{1.5}
\begin{tabular}{|c||c|c||c|c|} \hline
~~$X$~~& ~~$\lambda_2^{(1)}$~~&~~$n^{(1)}_{\rm s}$~~& ~~$\lambda_2^{(2)}$~~&~~$n^{(2)}_{\rm s}$~~ \\ \hline
$2.4\times10^4$&$0.469\times10^{-7}$&$0.957$&$0.635\times10^{-7}$&$0.979$ \\ \hline
$2.2\times10^4$&$0.417\times10^{-7}$&$0.949$&$0.795\times10^{-7}$&$0.997$ \\ \hline
$2.0\times10^4$&$0.393\times10^{-7}$&$0.944$&$0.952\times10^{-7}$&$1.012$ \\ \hline
$1.8\times10^4$&$0.376\times10^{-7}$&$0.941$&$1.130\times10^{-7}$&$1.027$ \\ \hline
$1.4\times10^4$&$0.356\times10^{-7}$&$0.938$&$1.622\times10^{-7}$&$1.061$ \\ \hline
$1.0\times10^4$&$0.344\times10^{-7}$&$0.935$&$2.500\times10^{-7}$&$1.108$ \\ \hline
$1.0\times10^3$&$0.334\times10^{-7}$&$0.933$&$3.229\times10^{-6}$&$1.628$ \\ \hline
$1.0\times10^2$&$0.334\times10^{-7}$&$0.933$&$(3.454\times10^{-5})$&($3.135$) \\ \hline
$1.0\times10^1$&$0.334\times10^{-7}$&$0.933$&$(3.528\times10^{-4})$&($7.851$) \\ \hline
\end{tabular}
}
\end{center}
\caption{Sample parameters sets for the $\mathcal{N}=60$ case. The
 maximal value of $X$ to achieve $\sqrt{A_{\rm s}} = 4.69\times 10^{-5}$ is $X=2.461\times 10^4$ with $\lambda_2=0.538\times 10^{-7}$ and $n_{\rm s}=0.965$. The parenthesized values are written just for completeness.}
\label{60}
\end{table}

\end{document}